\documentclass[12pt,preprint]{mnras}
\usepackage{graphicx}	% Including figure files
\usepackage{amsmath}	% Advanced maths commands
\usepackage{amssymb}	% Extra maths symbols
\usepackage{multicol}        % Multi-column entries in tables
\usepackage{bm}		% Bold maths symbols, including upright Greek
\usepackage{pdflscape}	% Landscape pages

\usepackage[T1]{fontenc}
\usepackage{ae,aecompl}

\usepackage{newtxtext,newtxmath}
\title[New formation pathways for T$\dot{\rm Z}$Os, magnetars, BHs and SNe]{Stellar Core-Merger-Induced Collapse: New Formation Pathways for Black Holes, Thorne - $\dot{\rm Z}$ytkow objects, Magnetars and Superluminous Supernovae}

\author[ Ablimit et al. ]{Iminhaji Ablimit$^{1,4,}$\thanks{E-mail:
iminhaji@nao.cas.cn (I.A.)}, Philipp Podsiadlowski$^{2}$, Ryosuke Hirai$^{3}$, James Wicker$^{5}$\\
$^{1}$CAS Key Laboratory for Optical Astronomy, National Astronomical Observatories, Chinese Academy of Sciences, Beijing 100101,
China.\\
$^{2}$Department of Physics, University of Oxford, Keble Rd, Oxford OX1 3RH, UK\\
$^{3}$Monash Centre for Astrophysics, School of Physics and Astronomy, Monash University, Clayton, Victoria 3800, Australia\\
$^{4}$Department of Astronomy, Kyoto University, Kitashirakawa-Oiwake-cho, Sakyo-ku, Kyoto 606-8502, Japan\\
$^{5}$National Astronomical Observatories, Chinese Academy of Sciences, Beijing 100101,
China. }
\begin{document}
\label{firstpage}
\pagerange{\pageref{firstpage}--\pageref{lastpage}}
\maketitle

\begin{abstract}

Most neutron stars (NSs) and black holes (BHs) are believed to be the
final remnants in the evolution of massive stars.  In this study, we
propose a new formation channel for the formation of BHs and peculiar
NSs (specifically, magnetars and Thorne-$\dot{\rm Z}$ytkow objects
[T$\dot{\rm Z}$Os]), which we refer to as the core merger-induced
collapse (CMIC) model. This model involves the merger during a
common-envelope phase of an oxygen/neon/magnesium
composition white dwarf and the core of a hydrogen-rich or helium-rich
non-degenerate star, leading to the creation of peculiar new types of
objects. The results of binary population synthesis simulations show
that the CMIC channel could make important contributions to the
populations of (millisecond) pulsars, T$\dot{\rm Z}$Os, magnetars and
BHs.  The possibility of superluminous supernovae powered by
T$\dot{\rm Z}$Os, magnetars and BHs formed through the CMIC model is
also being investigated. Magnetars with immediate matter surroundings
formed after the CMIC might be good sources for fast radio bursts.

\end{abstract}


%%%%%%%%%%%%%%%%%%%%%%%%%%%%%%%%%%%%%%%%%%%%%%%%%%

%%%%%%%%%%%%%%%%% BODY OF PAPER %%%%%%%%%%%%%%%%%%

\section{Introduction}

The remnants of core-collapse supernovae (CCSNe) at the end of the
evolution of single hydrogen-rich massive stars ($\gtrsim 8\,M_\odot$)
can be neutron stars\footnote{NSs also may be born through the
  electron-capture SNe (ECSNe from $\sim 8-10 M_\odot$ main-sequence
  stars) (e.g., Miyaji et al. 1980).} (NSs) or stellar-mass black
holes (BHs) (e.g., Baade \& Zwicky 1934; Woosley et al. 2002; Heger et
al. 2003).  There are also other formation pathways to form NSs and
BHs, such as accretion-induced collapse (AIC) or merger-induced
collapse (MIC) in interacting binaries: in the AIC channel, a white
dwarf (WD) with an oxygen/neon/magnesium (ONeMg) composition or a NS
accretes matter from a companion, the donor star, and grow in
mass to a critical limiting mass (e.g., Michel 1987; Ivanova \& Taam
2004; Ablimit \& Li 2015) when they collapse to an even more compact state and become a NS
or BH, respectively (e.g., Nomoto \& Kondo 1991; Timmes et al. 1996).
The merger of two WDs or the merger of two NSs may form NSs or BHs in
the MIC pathway; these are also very promising sources for producing
gravitational waves (GW) (e.g., Ivanova et al. 2008; Bauswein et
al. 2013; Shapiro 2017; Lyutikov \& Toonen 2019; Ruiter et al. 2019; Chattopadhyay et al. 2020; Abbott et
al. 2020).  The AIC route to form NSs is an important channel to
produce pulsars and magnetars (e.g., Hurley et al. 2010; Tauris et al. 2013; Ablimit
\& Li 2015; Jones et al. 2016; Ablimit 2019, 2022), including young pulsars in old populations
(van den Heuvel 1981, 1987), and are promising contributors to the NS
population in globular clusters because of their expected low kick
velocities (e.g., Podsiadlowski et al. 2004; Ivanova et al. 2007).
BHs formed through AIC or MIC may become members of low-mass or
intermediate-mass X-ray binaries or even become single BHs (e.g.,
Belczynski \& Taam 2004; Bauswein et al. 2013; Bernuzzi et
al. 2020). The upper limit for the mass of NSs is a key parameter in
these formation channels. However, the upper mass limit of NSs, above
which they collapse to BHs, is still controversial due to poor
observational constraints (e.g., Margalit \& Metzger 2017).

In a binary, two stars can merge due to orbital angular momentum loss.
Stellar mergers in binaries are important for a variety of
astronomical phenomena, such as peculiar SNe, NSs and BHs. Indeed, a
significant fraction of single massive stars may be the product of the
merger of binary stars, and binary interactions, including mergers,
are important for producing various types of SNe and unusual objects
(e.g., Podsiadlowski et al. 1992; Langer 2012; de Mink et al. 2014;
Andrews et al. 2018; Schneider et al. 2019; Horiuchi et al. 2020; Mandel et al. 2021).
Wiktorowicz et al. (2019) studied possible merger routes to produce
BHs by performing binary population synthesis (BPS) simulations. In
the current study, we consider core mergers in WD binaries during the
common envelope (CE) phase as an alternative pathway to form unusual
objects including BHs.

Two stars evolve into a CE if mass transfer by Roche-lobe overflow in
a binary is dynamically unstable (Paczy$\acute{\rm n}$ski 1976).
Although CE evolution is still very poorly understood and not all
binaries undergo the CE (e.g., Kemp et al. 2021), it has been
seen as a crucial phase in binary evolution, irrespective of whether
the stars merge in the CE or survive from the CE (see Ivanova et
al. (2013) for a detailed review of the uncertainties and the
importance of this phase).  It has been suggested that future
observations of planetary nebulae and luminous red novae may provide
useful constraints to improve our understanding of many of the open
questions of the CE phase (Blagorodnova et al. 2017; MacLeod et
al. 2018; Jones 2020). Kruckow et al. (2021) presented a catalog of
candidate post-CE binaries, and they showed that their
catalog and its future extensions will provide insight into the
evolution of close binaries through the CE phase. As long as a few decades ago, it has been
suggested that the merger inside a CE could potentially trigger a
peculiar supernova event (Sparks \& Stecher 1974).  The merger of a CO
(carbon/oxygen) WD and the core of an asymptotic giant branch (AGB)
star during the CE phase has been proposed as a possible progenitor
channel for Type Ia SNe (e.g., Soker 2011; Soker et al. 2013). Mergers
between WDs and cores of hydrogen-rich normal stars during the CE
phase have been pointed out for the possible origins of
intermediate-luminosity optical transients (Red Transient; Red Nova)
and/or peculiar SNe (Sabach \& Soker 2014).  More recently, Ablimit
(2021) investigated stellar core mergers between CO WDs and cores of
hydrogen-rich intermediate-mass, cores of massive stars, or cores of
helium-rich non-degenerate stars during the CE phase as origins for
peculiar type Ia SNe and type II SNe with the framework of the
core-merger detonation (CMD) model. The Galactic rates calculated with
the CMD model show that this scenario has significant contributions to
SNe Ia (i.e. overluminous/super-Chandrasekhar-mass SNe Ia and SNe Iax)
and SN 2006gy-like type II superluminous SNe (SLSNe; see Ablimit (2021)
for the details of CMD model and discussions).

The merger of a NS and the core of massive companion star
  inside the CE can produce a Thorne-$\dot{\rm Z}$ytkow object (T$\dot{\rm Z}$O;
  Thorne \& $\dot{\rm Z}$ytkow 1977; Taam et al. 1978).  T$\dot{\rm Z}$Os are an
  exotic class of stellar objects comprised of a neutron core
  surrounded by a extended envelope; from the outside they have the appearance of
  a red supergiant (Thorne \& $\dot{\rm Z}$ytkow 1977;
  Levesque et al. 2014; DeMarchi et al. 2021).  They might eventually
  explode as (SLSNe)\footnote{Superluminous SNe (SLSNe) have
    a luminosity around $10^{44}\,\rm erg\,\rm s^{-1}$ and radiate a
    total energy of $\sim 10^{51}\,\rm erg$ over a period of several
    months (see Moriya et al. (2018a) and Gal-Yam (2019) for
    reviews). Type I SLSNe do not show hydrogen lines in their
    spectra, while Type II SLSNe do.} (Moriya 2018; also see Chevalier
  (2012)\footnote{In Chevalier's paper, the SLSN occurs in the
    spiral-in process itself, not in a T$\dot{\rm Z}$O.}).  Magnetars and
accreting BHs have also been studied as potential central engines for
powering SLSNe (Kasen \& Bildsten 2010; Woosley 2010; Inserra et al.
2013; Dexter \& Kasen 2013). In this study we consider a possibly
important (see our results and discussions for important contributions
of the channel to peculiar transients and objects),
alternative channel for producing SLSNe, involving T$\dot{\rm Z}$Os,
magnetars and BHs.

%Mergers of CO WDs with cores of stripped helium stars inside the CE are also studied as
%possible progenitors for peculiar SNe (Maeda \& Ablimit 2020; Ablimit et al. 2020).
%Recently, the merger between a CO WD and He core of a massive giant star has been suggested to reproduce the observational constraints
%placed for SN 2006gy, and this merger scenario has been investigated as progenitors for SN 2006gy-like type II superlumious SNe
%(Jerkstrand et al. 2020; Maeda \& Ablimit 2020; Ablimit et al. 2020).
%Zapartas et al.(2019) showed that binary mergers dominate the binary channels leading to type II core-collapse SNe,

Here, we systematically explore the presumed core-collapse SN-like event
following the merger of an ONeMg WD and the core of a non-degenerate
star inside a CE.  We call this channel, involving the merger of a
ONeMg WD with the He core of a massive star in a CE, the core
merger-induced collapse (CMIC) scenario in order to distinguish it
from the more `normal' AIC because the timescale of CMIC is much
shorter than the timescale of AIC in accreting ONeMg WDs through
stable mass transfer. We propose that the CMIC scenario could be an
alternative formation channel for single (millisecond) pulsars,
magnetars and BHs with an accretion disk, and that SLSNe could
be powered by these compact objects formed through CMIC.
We also study the mergers between ONeMg WDs and cores of stripped,
non-degenerate helium stars and low-/intermediate-mass stars as a
potential channel to produce single (millisecond) pulsars. The rates
of ONeMg WD -- NS binaries surviving the CE phase are also being
investigated as they are potential GW sources.  We also explore how
the final outcomes of these binary populations depend on the modeling
of the more poorly understood physical phases and key parameters in
binary evolution studies, specifically involving mass transfer (e.g.,
its stability depending on the mass ratio and the mass-transfer
efficiency), the modeling of the common-envelope phase (e.g, the CE
efficiency and the binding energy parameter), metallicity, etc.  In \S
2, we describe our method to treat the physical binary processes in
the BPS simulations and discuss the assumptions for the CMIC.  The
main results are presented in \S 3. The possible contribution and
possible observational implications are discussed in \S 4, and a
summary is given in \S 5.

\section{Methods}
\label{sec:model}

\subsection{Binary Model Set}

In order to derive the population of WD$+$non-degenerate star
binaries, we perform Monte-Carlo simulations by using an updated
$BSE$ population synthesis code based on Hurley et al. (2002; see Ablimit
et al. (2016) and Ablimit \& Maeda (2018) for the updated version), starting with $10^7$
primordial binaries (consisting of two zero-age main sequence stars) with chosen
distributions of primary mass, secondary mass and orbital
separation.
Initial binary parameters of primary mass, secondary mass and orbital
separation can be set up with the $\rm n_x$ grid points of parameter $\chi$ logarithmically spaced,
\begin{equation}
\delta {\rm ln\chi} = \frac{1}{\rm n_x - 1}({\rm ln{\chi_{\rm max}}} - {\rm ln{\chi_{\rm min}}}),
\end{equation}
For each set of initial parameters, we evolve the binary system to
an age of the Hubble time, or until it is destroyed. Each phase of the
evolution, such as tidal evolution, angular momentum changes due to mass variations etc.,
is followed in detail according to the algorithms described in Hurley et al.(2002).
The initial input parameters of the primordial binaries are set the same as in
Ablimit et al.(2016).  The initial mass function of Kroupa et
al. (1993) is adopted for the primary star ($M_1$),
\begin{equation}
f(M_1) = \left\{ \begin{array}{ll}
0 & \textrm{${M_1/M_\odot} < 0.1$}\\
0.29056{(M_1/M_\odot)}^{-1.3} & \textrm{$0.1\leq {M_1/M_\odot} < 0.5$}\\
0.1557{(M_1/M_\odot)}^{-2.2} & \textrm{$0.5\leq {M_1/M_\odot} < 1.0$}\\
0.1557{(M_1/M_\odot)}^{-2.7} & \textrm{$1.0\leq {M_1/M_\odot} \leq 150$},
\end{array} \right.
\end{equation}
The distribution of the mass of
the secondary ($M_2$;
where the secondary is the star with the lower initial mass)
is determined by the distribution of the initial mass ratio,
\begin{equation}
n(q) = \left\{ \begin{array}{ll}
0 & \textrm{$q>1$}\\
\mu q^{\nu} & \textrm{$0\leq q < 1$},
\end{array} \right.
\end{equation}
where $q=M_2/M_1$ and $\mu$ is the normalization factor for the
assumed power-law distribution with index $\nu$. We consider a flat
distribution ($\nu = 0$ and $n(q)=$constant) for the initial
mass-ratio distribution. The chosen distribution for the initial orbital
separation, $a_{\rm i}$, is
\begin{equation}
n(a_{\rm i}) = \left\{ \begin{array}{ll}
0 & \textrm{$a_{\rm i}/R_\odot < 3$ or $a_{\rm i}/R_\odot > 10^{6}$},\\
0.078636{(a_{\rm i}/R_\odot)}^{-1} & \textrm{$3\leq a_{\rm i}/R_\odot \leq 10^{6}$} \ .
\end{array} \right.
\end{equation}
We also adopt a thermal distribution for the initial eccentricity
between 0 and 1.  For the SN remnant calculation, we adopted the rapid
remnant-mass model of Fryer et al. (2012) in the subroutine $hrdiag$
of the updated $BSE$ code, and note that a typo in Fryer et
al. (2012) has been corrected in our code (corrected as $a_1 = 0.25 -
1.275/{(M_1 - M_{\rm proto})}$, and the proto-compact object
mass is set as $M_{\rm proto}=1.0 M_\odot$). For the natal supernova kick
velocity, imparted to the newborn NS at its birth, we adopt Maxwellian
distributions with a velocity dispersion of $\sigma_{\rm k} =
265\,{\rm km\,s^{-1}}$ (Hobbs et al. 2005) for CCSNe and $\sigma_{\rm
  k} = 40\,{\rm km\,s^{-1}}$ for ECSNe/AIC (e.g., Hurley et al. 2010).
  For the wind mass loss,
the prescription of Vink et al. (2001) is used for O and B stars in
different stages (hot stars), and $1.5\times10^{-4}\,\dot{M}_\odot\,\rm
yr^{-1}$ for luminous blue variables (Vink \& de Koter 2002) in the
subroutine $mlwind$ of the code (see the wind model 2 of Ablimit \&
Maeda (2018)). Note that these prescriptions (for massive
  stars), i.e., the remnant-mass determination, wind mass loss and
kick velocity do not influence the main results of this study as the
formation of the ONeMg WD systems of interest here never suffer from
CCSN explosions nor do they involve very massive stars. One of
  the important issues is how ONeMg WD form. Stellar evolution
  remnants composed mainly of O and Ne after the ejection of the
  envelope of a thermally pulsing AGB stars are considered as ONeMg
  WDs in the code (e.g., Hurley et al. 2000). Note that the mass range
  of stars that form ONeMg WDs strongly depends on the modeling of a
  variety of, sometimes uncertain, processes in stellar evolution
  codes, such as the initial composition, the adopted overshooting,
  nuclear reaction rates and the role of rotation (e.g., Doherty et
  al. 2017).

As discussed in section 1, CE evolution is a key phase in the
evolution of many binary systems. Whether it occurs depends most
importantly on the mass ratio.  If the mass ratio is larger than a
critical mass ratio ($q_{\rm c}$), mass transfer between the two
binary components is dynamically unstable and a CE forms.  When the
donor star is on the main sequence (MS) or crosses the Hertzsprung gap
(HG), our adopted default value is $q_{\rm c} = q_{\rm const} = 4.0$
(Hurley et al. 2002). As an
alternative we also use a prescription of $q_{\rm c} = q_{\rm cs}$ in the subroutine $evolv2$
of the code, where $q_{\rm cs} = M_1/M_2$ is determined with the mass-transfer model by taking the spin of the
  accretor into account (for more details see Ablimit et al. (2016) and Ablimit \&
Maeda (2018), and see Table 2). This rotation-dependent mass accretion model leads
to the accretion efficiency, because the accretion rate of a rotating accretor is reduced by a
  factor of ($1 -\Omega / {{\Omega}_{\rm cr}}$), where ${\Omega}$ is
  the angular velocity of the star and ${\Omega}_{\rm cr}$ is its
  critical rotation value (e.g., Stancliffe \& Eldridge (2009)).  An
  accreting star can spin very rapidly even when only a small amount
  of mass has been transferred from the donor (Packet 1981); this
  drastically reduces the mass-transfer efficiency (to values $< 0.2$)
  and can even stop it completely when the accretor rotates at
  ${\Omega}_{\rm cr}$ (but also see Paczy$\acute{\rm n}$ski (1991)).
  The mass-transfer efficiency determines how much of the transferred
  matter is accreted by the accretor and how much escapes from the
  binary. It is assumed that the matter is escaped/ejected from
  the binary when it is no longer affected by the binary.
  We assume that material that is ejected from the system
  carries the specific angular momentum of the accretor (e.g., Hurley
  et al. 2002).  In this prescription, the maximum initial mass ratio
  of the primary to the secondary in the primordial binaries can be as
  high as $\sim$5-6, and a larger number of primordial binaries can
  avoid a contact phase and hence experience stable mass transfer
  until the primary's envelope is completely exhausted (see Abdusalam
  et al. (2020) for similar discussions and results).  When the
original primary is on the first giant branch (FGB), AGB or is a
helium (He) star, the prescriptions in Hurley et al. (2002) are
applied.

The standard CE model,  based on energy
conservation (Paczy$\acute{\rm n}$ski 1976; Ivanova 2013), is utilized in our simulations,
\begin{equation}
E_{\rm{bind}} = {\alpha_{\rm CE}} \Delta E_{\rm orb} \ ,
\end{equation}
where $E_{\rm{bind}}$ and $\Delta E_{\rm orb}$ are the binding energy
of the envelope and the change in the orbital energy during the CE
phase, respectively. In the specific model of Webbink (1984), which we
adopt here, the binding energy of the envelope is parameterized as,
\begin{equation}
E_{\rm{bind}} = - \frac{GM_1 M_{\rm{en}}}{\lambda {R}_1},
\end{equation}
where $M_1$, $M_{\rm{en}}$ and ${R}_1$ are the total mass, envelope
mass and radius of the primary star, respectively.  We use two values
for the CE efficiency as $\alpha_{\rm CE} = 1.0$ and $\alpha_{\rm CE}
= 0.1$ in the subroutine $comenv$ of the code. The value of
  $\lambda$ varies when a star evolves and strongly depends on the
  star's initial mass, its evolutionary state and other possible
  energy sources, such as, e.g., recombination energy (see, e.g., Han
  et al.\ 1994; Tauris \& Dewi 2001; Podsiadlowski et al. 2003; Wang
  et al.  2016a,b). The results of Wang et al. (2016a,b) are supported by
   Klencki et al. (2021), they are also consistent with results of the very
   recent CE study of Vigna-G$\acute{\rm o}$mez et al. (2021), and these consistency show
   the fact that results of Wang et al. (2016a,b) are the correct and reliable ones. We take either a
  constant value $\lambda = 0.5$ or the detailed prescription $\lambda
  = \lambda_{\rm w}$ ($\lambda_{\rm w}$ is taken from $\lambda_{\rm b}$ of Wang et al. (2016b)) for the binding energy
  parameter. Whether the two stars merge completely or survive from
  the CE to continue their evolution critically depends on these two
  parameters. We set the initial metallicity of stars to be $Z=0.02$ and
  $Z=0.001$.  Table 2 summarizes our main simulation models. For other
  parameters we use the default values in Hurley et al. (2000,
  2002).

\subsection{The Core Merger-Induced Collapse during the CE phase}

In the traditional AIC scenario, the ONeMg WD grows towards the
Chandrasekhar mass by accreting H- and/or He-rich material transferred
from its non-degenerate companion through stable mass transfer. In
addition to the AIC channel, two WDs in a compact binary (with a
combined mass $\geq$ the Chandrasekhar mass where at least one of them
is an ONeMg WD) may merge and collapse to a NS in the so-called
merger-induced collapse (MIC).  Because the energy released by nuclear
fusion in a O+Ne deflagration that is expected to take place during
the merging process is not sufficient to cause an explosion of the
tightly bound core (Miyaji et al. 1980), further electron captures
eventually lead to the gravitational collapse of the core and the
formation of a NS (e.g., Nomoto \& Kondo 1991).  AIC/MIC is the
favored alternative way to produce pulsars in globular clusters that
have characteristic ages significantly less than the age of the
clusters (Lyne et al. 1996; Boyles et al. 2011), and also to form
millisecond pulsars (MSPs) in addition to the standard recycling
scenario involving NSs formed in a CCSN (Chanmugam \& Brecher 1987;
Michel 1987; Kulkarni \& Narayan 1988; Bailyn \& Grindlay 1990).

In this work we consider the case where, after a ONeMg WD CE system
formed, the two cores, i.e. the WD and the core of the donor star,
spiral towards each other; if the CE cannot be expelled, they
eventually coalesce and mix completely when the cores are both
degenerate. If one core is considerably more compact (usually a WD)
than the other (e.g. a non-degenerate He core), the non-degenerate
component will eventually be tidally disrupted.  The WD will sink to
the center of the CE system without mixing, while it rapidly accretes
material from the disrupted component.  In both cases, the final
outcome will be the collapse of the WD to form a more compact object.
We present several evolutionary routes involving a CE phase, assuming
that the CMIC happens inside the CE if the combined mass of an ONeMg
WD and the core mass of the companion star is larger than the
Chandrasekhar limit mass (1.44$M_\odot$).

How the final merging of the compact core and the companion object
happen is not very well understood, and it is not entirely clear
whether and how the core is being spun up, but there are various
possibilities for spin-up. For example, the matter stream of the
spiraling-in companion may impact with the more compact core and
spin it up in a "slow merger" phase (e.g., Podsiadlowski et al. 2010).
Another possibility is that the spiral-in occurs on a dynamical
timescale and the companion is immediately dynamically disrupted; this
would naturally lead to some type of accretion disk around the more
compact component from which it could be spun-up by disk accretion.

If the combined core masses are larger than 1.44$M_\odot$ and the
  total mass (the two cores and the envelope) is less than 15
  $M_\odot$ during the CE, the core merger through the CMIC could form
  NSs/pulsars (of which a fraction could be MSPs) or magnetars (see
  Figure 1 as one example; this will be further discussed in
  subsequent sections).  The newborn NS formed by CMIC is likely to be
  surrounded by a substantial amount of material because not all the
  CE material may be ejected from the newborn NS in the collapse
  event\footnote{For comparison, only 10\% of the envelope typically
    becomes unbound when MS stars merge (e.g., Owocki et al. 2019), and
    the unbound ejecta is probably around 1\% for degenerate mergers
    (e.g., see Fig. 13 in Dan et al. (2014)).}.  Here, the newborn NS
  formed from CMIC in the WD + massive star binary could be an unusual
  type of object, because this NS would be surrounded by a very
  massive envelope, potentially much more massive than the central compact
  object itself. Thus, this would produce a configuration similar to
  that of a T$\dot{\rm Z}$O (e.g., DeMarchi et al. 2021); we therefore postulate
  that T$\dot{\rm Z}$Os could be formed in the CMIC scenario.

As the envelope loses its thermal energy by radiation from the
  surface of the merger product, an accretion disk may form around the
  compact core, in particular as neutrino cooling could be very
  effective near the NS (e.g., Fryer et al. 1996; Taylor et al. 2011).
  Fryer et al. (1996) predicted that a range of rapid-infall neutron
  star accretors present in certain low-mass X-ray binaries,
  common-envelope systems, cases of supernova fallback, and T$\dot{\rm Z}$Os
  would lead to explosions by neutrino heating similar to the
  neutrino-driven explosions in standard CCSNe.

A certain fraction of known WDs in binaries has been found with
bipolar magnetic fields (Kahabka 1995; Sokoloski \& Bildsten 1999;
Osborne et al. 2001; Ferrario et al. 2015; Pala et al. 2020), and some
observed WDs are highly magnetized (e.g., Schmidt et al. 1999). The
magnetic field of a WD has been pointed out as one important physical
parameter that influences the evolution of a WD in a binary (e.g.,
Wheeler 2012; Ablimit et al. 2014; Farihi et al. 2018). Recently, detailed
and systematic binary evolution studies show that the magnetic field
of the WD can affect the accretion phase (Ablimit \& Maeda 2019a, b),
and Ablimit (2019) showed that it can play a crucial role in the AIC
formation route for peculiar MSPs. A magnetic WD can be
  spun up by gaining angular momentum during the merger process inside
  the CE as discussed above. If the WD in the CMIC channel is strongly
  magnetized, the newborn MSP/pulsar would inherit the large magnetic
  field through magnetic field amplification during the CMIC process
  of the highly magnetized WD (e.g., Duncan \& Thompson 1992). Indeed,
  several previous studies proposed that the large magnetic
  fields of magnetic WDs can be formed or amplified during CE
  evolution (e.g., Tout et al. 2008; Wickramasinghe et al. 2014;
  Ohlmann et al. 2016). Based on these observational and theoretical
  studies, we here hypothesize that a magnetar could be the result of
  this CMIC formation channel if it formed in a WD CE system with a
  highly magnetized WD.

The first magnetar was identified about three decades ago; typically
magnetars are isolated and relatively rapidly rotating neutron stars
(NSs) with very strong magnetic fields (up to $\sim 10^{15}$ G; Duncan
\& Thompson 1992). The majority of the magnetars known to date (29 in
the compilation of Olausen \& Kaspi (2014)) have been discovered just
over the past two decades (with a range of spin periods between 1 and
11\,s) through their distinctive high-energy phenomenology: bursts of
X-ray/gamma-ray emission and/or enhancements of their persistent X-ray
luminosity, dubbed 'outbursts' (see Kaspi \& Beloborodov 2017;
Esposito et al. 2021). A magnetar surrounded by ejected material in
this channel may power X-ray emission and possibly 'outbursts'.  It is
also possible that a BH may be formed in a CMIC during the CE phase
when the combined core masses are larger than 1.44$M_\odot$ and the
total mass (cores and envelope) is larger than 15 $M_\odot$; the reason
why we set 15 $M_\odot$ as a cut to form pulsars/magnetars or BHs will
be discussed further in subsections below (especially in subsection 3.2).
The possibility that
  the central newborn objects, such as T$\dot{\rm Z}$Os/magnetars/BHs,
  may power transient events like SLSNe will also be considered in
  later sections.

\section{Results}

We first examine the evolution of two MS stars leading to a CE event
involving an ONeMg WD and the core of a non-degenerate star.  In a
Monte-Carlo simulation of $10^7$ binaries with the standard model
parameters (model 1), there are 3535 CE events which involve ONeMg WDs
and non-degenerate stars, and 1907 of them end up with mergers between
the ONeMg WDs and the cores of the non-degenerate stars inside the CE.
We focus on the core mergers during the CE phase as progenitors of some
transients and objects and considering different binary stellar physics
assumptions to estimate the importance of the different channels.

\subsection{Pulsars, Thorne - $\dot{\rm Z}$ytkow object, Magnetar and Superluminous SNe through CMIC inside the CE }

In a typical example, illustrated in Figure 1, the primary star
evolves through the MS, HG, helium-rich MS (HeMS) and helium-rich
giant (HeG) phases and forms an ONeMg WD\footnote{Note that the mass of zero-age MSs to
form ONeMg WDs may be different (comparing to single stellar evolution) due to
possible early mass loss or accretion caused by the binary interaction (e.g., Hurley et al. 2000, 2002).}.
Because of the two stable
RLOF mass-transfer phases from the primary, the secondary becomes a
relatively massive star. The WD is engulfed by the mass gained from
the massive secondary when the secondary fills its Roche lobe during
its core helium burning (CHeB) stage.  The system fails to eject the
CE, and the WD and the core of the massive secondary merge during the
CE phase.  This evolutionary route of the merger between an ONeMg WD
and the core of a H-rich normal star finally produces a NS. The large
amount of angular momentum transferred onto the compact object during
the merger process may make the newborn NS a fast rotator like a pulsar
or a millisecond pulsar.  If the WD has a strong magnetic field before
the merger, the newborn NS most likely would become a magnetar.

The typical ranges of the initial binary parameters for the ONeMg WD +
H-rich star CE event channel are: initial primary masses, $M_{1,\rm
  i}\sim6.3-11.0M_\odot$, secondary initial masses, $M_{2,\rm
  i}\sim1.0-10.2M_\odot$; initial orbital periods, $P_{\rm orb,
  i}\sim0.01-122.0$ days. Figure 2 shows the mass distributions
  in model 1 for the merger case which cause CMIC.  Compared
  to the initial binary parameters of all ONeMg WD + H-rich star systems
  with a CE, the initial conditions (mass ranges) of binaries that lead to
  the core merger during the CE are quite similar (Figure
  2).  Figure 2 shows that mergers with more massive
  donors generally have a larger CE mass.

The rates of all ONeMg + non-degenerate star systems
  associated with a CE and core merger events for the different models
  considered are summarized in Table 3.  Table 3 shows that the rates
  of all ONeMg WD + non-degenerate H-rich star CE events vary
  significantly with the $\alpha$ parameter, metallicity and mass
  transfer assumptions.
  The pessimistic value of $\alpha=0.1$ in
model 2 causes more mergers between non-degenerate stars during the CE
events, therefore the number of ONeMg WD binaries with CE phases are
reduced from 0.33\% to 0.15\%. The model with the pessimistic value
$\alpha=0.1$ is adopted for illustrating the effect of $\alpha$,
though it should be noted that with such a pessimistic value, no close
double NS systems as GW sources can be formed (see Table
3). For all merger cases (0.005\% - 0.20\% mergers of $10^7$
  binaries), more binaries can have stable mass transfer and avoid the
  CE phase with the mass-transfer (critical mass ratio) model adopted
  in model 5 (see Section 2), thus reducing CE events in general.
  Because more energy (i.e. internal energy) is considered in the
  $\lambda$ prescriptions of models 4 and 5 for the CE expulsion,
  there are more successful CE ejections in the binaries, which
  reduces the number of CE merger systems. Zapartas et al. (2017)
  studied mergers of ONeMg WDs and HG stars as a contributor to
  'late' core-collapse SNe, and their rate ($\sim 2.1\times10^{-4}{M_\odot}^{-1}$)
  is in good agreement with our rate derived with mergers of ONeMg WDs
  and normal stars ($\sim 1.9\times10^{-4}{M_\odot}^{-1}$;
  see R1 or R3 in line 2 of Table 3). In this work,
  we investigated more new pathways as origins of various
  transients and peculiar objects comparing to literatures.

Now first consider cases with condition (1), i.e., systems
  with a combined core mass $M_{\rm com}\geq1.44 M_\odot$ and total
  mass $M_{\rm all}< 8.0M_\odot$ for mergers during the CE phase. This
  includes mergers between ONeMg WDs and He cores of evolved
  hydrogen-rich stars, and also mergers between ONeMg WDs and
  degenerate cores during the late stages of AGB stars (e.g., Canals et al. 2018).  The evolutionary
  delay times in this channel range from $\sim 0.1$ to 12.5
  Gyr. The rates for these systems range from
  $0.24\times10^{-5}{M_\odot}^{-1}$ to
  $7.38\times10^{-5}{M_\odot}^{-1}$ for the different models in Table 3.
  The newborn NSs with condition (1) could be MSPs with
  relatively smaller envelope mass (see Section 2 for the discussion
  of formation pathway, and Figure 2).
  Derived rates with CMIC are consistent with rates of AIC
  in previous studies (e.g., Hurley et al. 2010; Ablimit 2022).

Mergers with condition (2) (i.e.\ $M_{\rm com}\geq1.44
  M_\odot$ and $8.0\leq M_{\rm all}< 15.0M_\odot$) tend to form NSs
  surrounded by a large amount of material because of the more massive
  CE (see the discussion in Figure 2).  As discussed in Section 2,
  newborn NSs surrounded by massive envelopes naturally produce
  T$\dot{\rm Z}$Os. The typical T$\dot{\rm Z}$O formation rate from
  the CMIC channel is $\sim 6\times10^{-5}\,{\rm yr}^{-1}$ if we adopt
  a star-formation rate of 2 $M_\odot \, {\rm yr}^{-1}$ (e.g.,
  Misiriotis et al. 2006; Robitaille \& Whitney 2010) and binary
  fraction of 0.7; our rate is lower than the rate of $\sim
  2\times10^{-4}\,{\rm yr}^{-1}$ estimated by Podsiadlowski et
  al. (1995).  The evolutionary delay times in this channel are
between $\sim 14.18$ and 86.22 Myr.  If we consider systems with a
total mass $M_{\rm all}\geq 10.0M_\odot$ for producing T$\dot{\rm
  Z}$Os, the Galactic rate from the CMIC channel would be one order of
magnitude lower than the estimate of Podsiadlowski et al. (1995), and
these T$\dot{\rm Z}$Os may evolve into other objects on a relatively
short timescale. T$\dot{\rm Z}$Os born here would have very
  massive envelope (often larger than 8$M_\odot$).  Moriya (2018) has
  studied the strong accretion that occurs onto the central neutron
  core when the nuclear reactions that support such massive T$\dot{\rm
    Z}$O terminate. As the accretion is neutrino driven, it can be
  highly super-Eddington and trigger a supernova event when the NS
  contracts and ultimately collapses. Such explosions would be
  observed as energetic Type II supernovae or even as superluminous
  supernovae of Type II.  A strong large-scale outflow or a jet can be
  launched from the super-Eddington accretion disc formed during the
  collapse (Moriya 2018).  The fallback of material towards the
  neutron star after a successful explosion is large in this case, and
  a black hole is formed very quickly (i.e. a few seconds; Fryer et
  al. 1996).

With conditions (1) and (2), it is possible that the newborn
  NSs/pulsars/MSPs formed through CMIC could be magnetars if the
  collapsing WDs are strongly magnetized (as discussed in section 2).
  The fraction of highly magnetized WDs among the WD population
  has been estimated from observations to be around 15\%
  ($\sim$15\%-18\%; Ferrario et al. 2015; see also Kepler et al. 2013,
  2015).  Adopting 15\%, we can estimate a possible birthrate for
  magnetars via CMIC. It lies between
  $0.03\times10^{-5}{M_\odot}^{-1}$ and
  $2.46\times10^{-5}{M_\odot}^{-1}$ for the different chosen models (see Table
  3).  Note that these rates are simply obtained by combining the results for
  conditions (1) and (2) and multiplying them with 0.15.
  These rates have a good agreement with the Galactic rate of magnetars formed
  by AIC from magnetized WD binaries (Ablimit 2022). A rotating
  magnetar radiating according to the classic dipole formula could
  power a very luminous supernova (e.g., Woosley 2010); thus magnetars
  formed via CMIC may power Type II SLSNe. We will compare the
  predicted rates with observationally inferred rate of peculiar SNe
  in Section 4.

The combined total mass distributions (Figure 3) for mergers
demonstrate that metallicity, the details of mass transfer and the binding energy
parameter in the CE phase play an important role. The
  central temperatures and densities are higher, and more hydrogen
  burns stably for the lower-metallicity and more massive MS stars;
  this leads to larger core masses and different evolutionary lifetimes
  for the merger systems. Thus, more ONeMg WD binaries
  can be formed at lower metallicity and for a higher total mass.
  There are few or no WD CE merger systems with massive
  star companions (especially with a combined mass $> 15{M_\odot}$) in models 4 and 5,
  in comparison to the other models, because more massive star systems can
  survive or avoid the CE phase for the adopted $\lambda$ and
  mass-transfer prescriptions in model 4 and 5 (Table 1 and Figure 3)
  for the same reasons as discussed above.

Now consider ONeMg WD CE events in which the companion star is a
non-degenerate He star.  The ONeMg WD + He star CE event channel is
realized for the following ranges in the initial binary parameters:
initial primary mass, $M_{1,\rm i}\sim6.34-9.1M_\odot$, secondary
initial mass $M_{2,\rm i}\sim2.1-7.9M_\odot$ and initial orbital
period $P_{\rm orb, i}\sim 0.03-40.0$ d.  The mass range of the ONeMg
WDs in this channel lies between $\sim 1.07$ and $\sim 1.42 M_\odot$
while the He stars' mass is mainly distributed between $\sim 0.75$ and
$\sim 1.5 M_\odot$ (up to $\sim 2.25 M_\odot$; see Figure
4). The rates ($0.34 - 10.78\times10^{-5}{M_\odot}^{-1}$) in
  Table 2 show that at most $\sim$10\% of all CE events of CO WDs $+$
  He stars end up with mergers for the condition $M_{\rm com} \geq
  1.44M_\odot$.  The evolutionary delay times lie between $\sim$0.05
  and 1.0 Gyr.  The different patches in Figure 4 are caused by
  different evolutionary routes to form WD $+$ He star binaries.
  Figure 4 shows that the mass distributions for the cores and
  envelopes of He stars for the standard model 1 have a relatively
  narrow range between $\sim 0.3$ and $\sim 1.0 M_\odot$.  The lower
  rates and narrow mass ranges suggest that ONeMg WDs that merge with
  the core of a He star companion may not play as important a role as
  mergers with different types of companions. However, this might be
the origin for some rare transients and objects where no hydrogen
lines are detected. Our result suggests that this channel could
produce millisecond pulsars as well as magnetars.  Moriya et
al. (2018b) studied the fallback accretion central engine model for
hydrogen-poor SLSNe; they showed that $\geq 2M_\odot$ needs to be
accreted for powering SLSNe by fallback accretion.  The derived
envelope masses in Figure 4 are smaller than the minimum value
suggested by Moriya et al. (2018b; Nicholl et al. 2017); therefore it
is not clear whether the central objects formed in this channel may be
able to produce fallback accretion powered type I SLSNe.

\subsection{Possible BH Formation through CMIC inside the CE}

Stellar evolution models suggest that stellar mass BHs are the final
products of massive stars $\gtrsim 20-25\,M_\odot$ (e.g., Woosley \&
Weaver 1995).  Recent numerical simulations (e.g., O'Connor \& Ott
2011; Ugliano et al. 2012; Sukhbold \& Woosley 2014; Pejcha \&
Thompson 2015; Ertl et al. 2016; Sukhbold et al. 2016; Schneider et
al. 2021) indicate that the outcome of neutrino-driven explosions is
largely controlled by the final core structure of massive stars, and a
clear mass limit for a progenitor star to end its evolution as a BH is
hard to establish. The likelihood of a successful explosion or
  of BH formation is related to the compactness of the stellar core
  (e.g., O'Connor \& Ott 2011; Sukhbold et al. 2016; Raithel et
  al. 2018).  Observational and theoretical results suggest different
  progenitor mass ranges for failed SNe and BH formation (Smartt 2009,
  2015; Kochanek 2014; Schneider et al. 2021). Sukhbold et al. (2016)
  suggested that single zero-age MS stars with masses as low as 15$M_\odot$ could
  produce BHs, while Schneider et al. (2021) showed that, for stripped
  stars in binaries, a much higher initial mass is likely to be
  required.  On the other hand, in binary systems there are other
  ways, e.g., stellar mergers (Wiktorowicz et al. 2019) that could
  produce BHs.  Here, we propose the CMIC scenario as an additional
  channel for producing BHs which is the final outcome of the merger
  of an ONeMg WD and the core of its companion during the CE phase.
  Here we assume that a BH is formed when the total mass of the system
  is larger than 15$M_\odot$ if the merger occurs in a CE. The newborn
  BH formed through CMIC may be surrounded by an accretion disc formed
  by the ejected material as shown in the typical example illustrated
  in Figure 5. Note that the core of the merger system with the whole
  mass of $15 M_\odot$ in our model is already an ONeMg WD, thus the
  core density is significantly higher than that of a normal MS star with
  the same mass studied in Sukhbold et al. (2016). In this merger
process, there may be a quick transition from a NS to a BH (e.g.,
Moriya et al. 2016) due to the heavy core and the large amount of
fallback material after the collapse of the heavy ONeMG WD (in Figure
5, we only show the final possible outcome).

The results of our calculations in Figure 2 show that the initial mass
for the primaries can be as high as $\sim$11 $M_\odot$. For the
mergers, the secondary mass, secondary core mass, secondary envelope
mass can be as high as $\sim 21 M_\odot$, $\sim
6.5 M_\odot$ and $\sim 17 M_\odot$, respectively. The ONeMg WDs tend to be more massive
($>1.2M_\odot$).  The mass distributions are mainly affected by the
metallicity, details of mass transfer and CE evolution (i.e., Figure
3; see the earlier discussion). The rate for BH formation in
  the CMIC channel is as high as $3.48\times10^{-5}{M_\odot}^{-1}$
  (Table 2). This implies that this channel makes a contribution of at
  least a few percent to the BH population. The massive CE would not
  be ejected far away during the CMIC and fast NS to BH transition
  process, and at least some fraction of the ejected material in a
  core collapse supernova explosion may remain bound to the compact
  remnant.  Thus, the BH formed in this way may be surrounded by an
  accretion disk due to the turn-around and fallback as shown in the
  example in Figure 5.  The accretion power released when material
  falls back onto the compact remnant at late times could power unusual
  SNe such as superluminous or otherwise peculiar supernovae (e.g.,
  Dexter \& Kasen 2013). Therefore, the BH formed in this proposed
  formation channel could potentially power type II SLSNe.

\section{Discussion}

$Rates/Contributions$: Adopting a star formation rate of 2
  $M_\odot \, {\rm yr}^{-1}$ and binary fraction of 0.7, we can
  estimate the Galactic rates of peculiar objects and transients
  formed via the CMIC scenario as follow (the two values given below
  refer to the standard model and our model 4, respectively):
  (1) MSPs/Pulsars have rates of $8.81\times10^{-5}\, {\rm yr}^{-1}$
  and $6.57\times10^{-5}\,{\rm yr}^{-1}$. These pulsars can be the
  explanation for the retention of NSs in globular clusters. (2) The rate of T$\dot{\rm
    Z}$Os, including the possibility of Type II CCSNe/SLSNe and BHs
  powered by AIC of the cores, are $1.36\times10^{-4}\, {\rm yr}^{-1}$
  and $5.76\times10^{-5}{\rm yr}^{-1}$.  (3) Magnetars and SLSNe
  powered by these magnetars have rates of $3.36\times10^{-5}\, {\rm
    yr}^{-1}$ and $1.95\times10^{-5}{\rm yr}^{-1}$. (4) BHs and
  possible SLSNe have rates of $3.23\times10^{-5}\, {\rm yr}^{-1}$ and
  $0.11\times10^{-5}\,{\rm yr}^{-1}$.  Contributions of three objects
  (T$\dot{\rm Z}$O, magnetar and BH) to produce SLSNe lie between
  $7.82\times10^{-5}{\rm yr}^{-1}$ and $\sim 2\times 10^{-4}{\rm
    yr}^{-1}$. If we combine the rates for BHs formed right after CMIC
  and formed through AIC of core NSs in T$\dot{\rm Z}$Os, the Galactic
  rates of BHs range from $5.87\times10^{-5}{\rm yr}^{-1}$ to
  $1.68\times 10^{-4}{\rm yr}^{-1}$; this means that these BH
  formation pathways make a non-negligible contribution to the overall
  BH population.

Up to April 2014, according to the ATNF Pulsar Catalogue (Manchester
et al. 2005), 2016 pulsars with measured values of spin period and its
time derivative had been discovered.  Among these, 1847 were single
pulsars, a few percent millisecond pulsars, 55 of them in SN remnants,
and 29 single magnetars. Based on the CMIC model results, about 1000
-- 6500 pulsars can be formed, a large fraction of which are likely to
be millisecond pulsars, and about 30-100 could be magnetars (if we
take the observed fraction of magnetized WDs into consideration for
CMIC; Note that here it should be combined rates of all three channels of CMIC).
Thus, those observed numbers given above can be reproduced by the CMIC
channel proposed in this work. However, it is worth noting that the consistency between
our rates and observational results only shows the importance of the formation scenarios
in this work, and it does not mean that other possible formation ways could be excluded.
Most interestingly,
previous studies show the importance of the immediate surroundings of a magnetar for fast radio bursts (FRBs)
production in relativistic shock wave models (e.g., Khangulyan et al. 2021).
Magnetars formed through CMIC in this work can be immediately surrounded by envelope matters,
thus the newborn magnetars in this work may produce FRBs.  Compared to the NS population, the
CMIC channel makes a small contribution to the BH population.  Because
the rate of type II SLSNe is not observationally well-determined yet,
we roughly take the rate of type II SLSNe as $\sim 10^{-5}{\rm
  yr}^{-1}$ according to Quimby et al. (2013). Considering all cases
of NS/BH central engine models in our study, the predicted rates range
from $\sim 10^{-5}{\rm yr}^{-1}$ up to $\sim 2\times 10^{-4}{\rm
  yr}^{-1}$). Table 4 shows general rates of SLSNe which might be
originated from mergers in different types of WD binaries. For rates of various
transients and different objects formed through the stellar-core mergers from WD
binaries, see Ablimit (2021) and this paper. From the table, it can be seen
that different assumptions (especially different models for the CE evolution and
mass transfer) could change the rate by more than an order of magnitude, and it again shows
the important role of CE evolution, mass transfer and metallicity (see above
sections for more discussions). Derived rates in the merger
scenarios are compatible with the observational estimate, although
both of these estimates are inherently uncertain (see below for other
limitations).

$Related\,further\,evolution$: Above we investigated only mergers
inside a CE. But what about the systems that survive from the CE as WD
binaries? WD binaries are the one kind of X-ray binaries, and these X-ray binaries have
the connection to GW sources (e.g., Abdusalam et al. 2020). One particularly interesting final outcome of a ONeMg WD
with a massive star companion evolved from the CE is a WD-NS binary or possibly a WD-BH system
with a sufficiently close orbit to merge within a Hubble time. These
WD-NS and WD-BH binaries are excellent GW sources and may create peculiar
transients (Portegies Zwart \& Yungelson 1999; Paschalidis et al. 2004;
Sabach \& Soker 2014; Margalit \& Metzger 2016; Toonen et al. 2018; Abdusalam et al. 2020; Bobrick et al. 2021),
potentially observable with current and future transient
surveys such as LISA. Because the formation of close WD-BH system is very model-dependent,
there is no such system with the rapid remnant-mass model of Fryer et al. (2012), and ONeMg WD- NS systems
are the final outcomes survived from the CE evolution of ONeMg WD-massive star binaries in this study.
After CE ejection, the two stars move closer
towards each other due to various angular-momentum loss mechanisms
(i.e. GW radiation), and the resulting close ONeMg WD-NS systems will
merge within the Hubble time if they are sufficiently close after the
CE phase. Table 3 shows the rates of these ONeMg WD-NS binaries for
the different models, where the rate could be as high as
$4.95\times10^{-5}{M_\odot}^{-1}$. The rather
  non-conservative mass-transfer model adopted in this work makes more
  primordial binaries have stable mass transfer, reducing the number
  of CE events.  With the potentially more reliable treatment of
  $\lambda$ in model 4, more binaries can successfully survive from the
  CE phase if they evolve into a CE phase in the first place.
  Considering the formation of WD-NS systems and the contribution of
  mergers to these peculiar objects, our model 4 is the most favorable
  one compared to the other models.

$Limitations/Uncertainties$: It is worth to raise that the
  critical total mass of the WD core merged system to form a BH or a NS is very
  uncertain (it might be lower or higher than the adopted value (15$M_\odot$) in this work),
  we just adopted a possible value according to the previous
  work as discussed above, and this needs further investigations. The central NS objects (e.g., pulsars and T$\dot{\rm
    Z}$Os etc.) formed through CMIC could eventually become BHs if
    they accretes enough mass from fallback/surrounded materials to exceed the maximum NS mass. One of the other biggest uncertainties in
this work is the detailed nature of newborn NSs formed through CMIC; we
have discussed various possibilities, but this clearly needs further
investigation. Besides, there are many physical processes and
parameters in the binary evolution pathways which still remain
uncertain, specifically the CE phase, the existence of accretion discs
around the collapsed object, and the physical response at very high
mass-transfer rates. It is relatively easy to simulate all these
parameters in BPS simulations, i.e. the initial conditions,
prescriptions for stable/unstable mass transfer, treatments of common
envelope (CE) evolution etc.  The numbers of binaries, special objects
and merger systems are affected strongly by the treatment of all these
physical processes and parameters, most of all by the parameters for
the CE phase and the nature of the mass transfer.  In our BPS
simulations we used a range of acceptable/common ways by considering
new evolutionary models.  Future more detailed theoretical and
observational studies should provide more information and constraints
to improve our understanding.

%\subsection{Comparison and discussion}

\section{Conclusion}

We have proposed several possible new evolutionary pathways to form
(millisecond) pulsars, Thorne-$\dot{\rm Z}$ytkow objects, magnetars
and BHs, through core merger-induced collapse during the CE phase of
ONeMg WD binaries with hydrogen-rich intermediate-mass/massive stars
and stripped helium-rich stars. We have systematically investigated
the CMIC model from ONeMg binaries as origin for those objects, and
discussed possibilities of powering two types of SLSNe and FRBs etc. By comparing
our simulations with observationally inferred values of known peculiar NS objects, BH population and
estimates of the SLSN rate, we conclude that the CMIC scenario can make
a significant contribution to those peculiar transients and objects. We expect that future
  observations with more information including robust estimates of
  the number/rate of these systems and transients will be useful for
  further testing the scenario proposed in this work.

\section*{Acknowledgements}

I.A. thank T. Moriya and S. Popov for useful comments.
This work is supported by NSF.

\section*{Data Availability:} The data underlying this article
will be shared on reasonable request to the corresponding author.

%%%%%%%%%%%%%%%%%%%%%%%%%%%%%%%%%%%%%%%%%%%%%%%%%%



%\bibliographystyle{mnras}


\clearpage

\begin{table}

%\begin{center}
\caption{ Meanings of abbreviations used in the text and
figures.}

\begin{tabular}{c c }
 \hline\hline
Abbreviation & Meaning \\
\hline
%\multicolumn{5}{l}{Chandrasekhar model with $Z=0.02$}\\
AGB & Asymptotic giant branch (star)\\
AIC & Accretion-induced collapse\\
BH & Black hole\\
BPS & Binary population synthesis\\
CCSN(e) & Core-collapse supernova(e)\\
CE & Common envelope\\
CHeB & Core helium burning (star)\\
CMIC & Core merger-induced collapse\\
CO & Carbon/oxygen\\
ECSN(e) & Electron-capture supernova(e)\\
GW & Gravitational wave\\
HeMS & Star on the equivalent of the main sequence\\
 & for hydrogen-poor helium-burning stars\\
HeG & Hydrogen-poor helium-burning giant\\
HG & Hertzsprung-gap (star)\\
k1 & Type of the primary star in a binary\\
k2 & Type of the secondary star in a binary\\
MIC & merger-induced collapse\\
MS & Main sequence\\
MSP & Millisecond pulsar\\
NS & Neutron star\\
ONeMg & Oxygen/neon/magnesium\\
SN(e) & Supernova(e) \\
SLSNe & Superluminous Supernovae\\
T$\dot{\rm Z}$O & Thorne-$\dot{\rm Z}$ytkow object\\
WD & White dwarf\\

\hline
\end{tabular}
%\end{center}
\end{table}

\clearpage

\begin{table}

%\begin{center}
\caption{Different models in our calculation }

\begin{tabular}{c c c c c c}
 \hline\hline
Model & $\alpha_{\rm CE}$ & $\lambda$ & $q_{\rm c}$ & $Z$ \\
\hline
%\multicolumn{5}{l}{Chandrasekhar model with $Z=0.02$}\\
 mod1 & 1.0 & 0.5 & $q_{\rm const}$ & 0.02 \\
 mod2 & 0.1 & 0.5 & $q_{\rm const}$  & 0.02 \\
 mod3 & 1.0 & 0.5 & $q_{\rm const}$ & 0.001 \\
 mod4 & 1.0 & $\lambda_{\rm w}$ & $q_{\rm const}$ & 0.02 \\
 mod5 & 1.0  & $\lambda_{\rm w}$ & $q_{\rm cs}$ & 0.02 \\

\hline
\end{tabular}
%\end{center}
\end{table}

\clearpage

\begin{table}

\begin{center}
\caption{Calculated rates ($R$) of ONeMg WD/non-degenerate star CE events for five different models (in $10^{-5}$ ${M_\odot}^{-1}$)}

\begin{tabular}{lccccccc}
 \hline\hline
CE events of ONeMg WDs/companion stars & $R1$ & $R2$& $R3$ & $R4$ & $R5$ \\

\hline
%\multicolumn{5}{l}{Chandrasekhar model with $Z=0.02$}\\
ONeMg WDs/normal stars (all CE) &33.04 & 15.43 & 41.34  & 33.94 & 15.94 \\
ONeMg WDs/normal stars (all merger) & 18.51 & 12.03 & 19.90  & 9.62 & 0.52 \\
ONeMg WDs/normal stars (merger con1) & 6.29 & 4.96 & 7.38  & 4.69 & 0.24\\
ONeMg WDs/normal stars (merger con2) & 9.75 & 4.57 & 9.04  & 4.12 & 0.00\\
ONeMg WDs/normal stars (merger con3) & 2.31 & 2.32 & 3.48  & 0.08& 0.00\\
\hline
ONeMg WDs/He stars (all CE) & 2.31 & 0.34 & 3.78  & 10.78  & 10.62\\
ONeMg WDs/He stars (Merger with $M_{\rm com}\geq1.44M_\odot$) & 0.56 & 0.26 & 0.36 & 1.06 & 0.47  \\
\hline
\hline
ONeMg WDs - neutron star systems (experienced CE) & 0.11 & 0.0 & 0.06 & 4.27 & 4.95  \\
\hline
\end{tabular}
\end{center}
Note: Rates in the table are calculated with $R=\frac{\rm Event\,numbers}{\rm Total\,mass(\sim 10^7M_\odot \,in\,this\,work)\,yielded\,in\,a\,simulation\,run}$. Normal stars mean sub-giant or giant branch stars with hydrogen-rich envelopes; He-rich stars are (sub)giant helium stars, and all mergers with He stars have combined core masses of $M_{\rm com}\geq1.44$.
Merger con1 means that the combined mass of the core and the secondary is $M_{\rm com}\geq1.44 M_\odot$ and the whole mass $M_{\rm all}< 8.0M_\odot$, merger con2 means that
$M_{\rm com}\geq1.44 M_\odot$ and $8.0\leq M_{\rm all}< 15.0M_\odot$,  and merger con3 that $M_{\rm com}\geq1.44 M_\odot$ and $M_{\rm all}\geq 15.0M_\odot$ during the CE phase. Event numbers can easily be calculated by multiplying the numbers in the table by $R\times10^7$, and the birthrate can be determined by choosing a star formation rate in $M_\odot\,{\rm yr}^{1}$ ($SFR$) and binary fraction ($f_{\rm bin}$) and multiplying the numbers in the table by $SFR\times f_{\rm bin}\times R$.
\end{table}

\clearpage

\begin{table}

\begin{center}
\caption{Galactic birthrates ($R_{\rm B}$) of stellar core-mergers which may produce SLSNe: a comparison between mergers of CO WD/massive star CE events ($R_{\rm B, S-CO}$) and mergers of ONeMg WD/normal star CE events with two massive cases ($R_{\rm B, S-ONe}$) for five different models (in $10^{-5}$ ${\rm yr}^{-1}$)}

\begin{tabular}{lcccc}
 \hline\hline
Model &$R_{\rm B, S-CO}$ ($M_{\rm WD}\geq 0.9M_\odot$) & $R_{\rm B, S-ONe}$ ($8.0\leq M_{\rm all}< 15.0M_\odot$)& $R_{\rm B, S-ONe}$
($M_{\rm all}\geq 15.0M_\odot$)  \\

\hline
%\multicolumn{5}{l}{Chandrasekhar model with $Z=0.02$}\\
mod1 &10.10 & 19.50 & 4.62  \\
mod2 & 24.36 & 9.14 & 4.64   \\
mod3 & 17.88 & 18.08 & 6.96  \\
mod4 & 7.08 & 8.24 & 0.16  \\
mod5 & 1.70 & 0.0 & 0.0  \\
\hline
\end{tabular}
\end{center}
Note: This table only shows the Galactic rates (adopting SFR as 2
  $M_\odot \, {\rm yr}^{-1}$) of SLSNe originated from mergers of between CO WDs ($M_{\rm WD}\geq 0.9M_\odot$) and cores of massive stars
($M_2 \geq 8.0M_\odot$) inside the CE, and from massive merger products ($M_{\rm all} \geq 8.0M_\odot$) of ONeMg WDs and non-degenerate hydrogen-rich stars CE events.
 This table does not include other possibilities like SLSNe triggered by mergers of between CO WDs ($M_{\rm WD}\geq 1.0M_\odot$) and cores of massive stars or mergers between WDs and cores of less massive ones ($< 8.0M_\odot$) or mergers of between WDs and cores of stripped He stars. For other transients like CMIC or SNe Ia or other possibilities for SLSNe, other tables including discussions in this work and Ablimit (2021) can be used, and different SFR can be adopted with rates of those tables in this work and Ablimit (2021) to calculate the Galactic rates for a real comparison with future works.
\end{table}

\clearpage

\begin{figure*}
\centering
\includegraphics[totalheight=3.7in,width=4.8in]{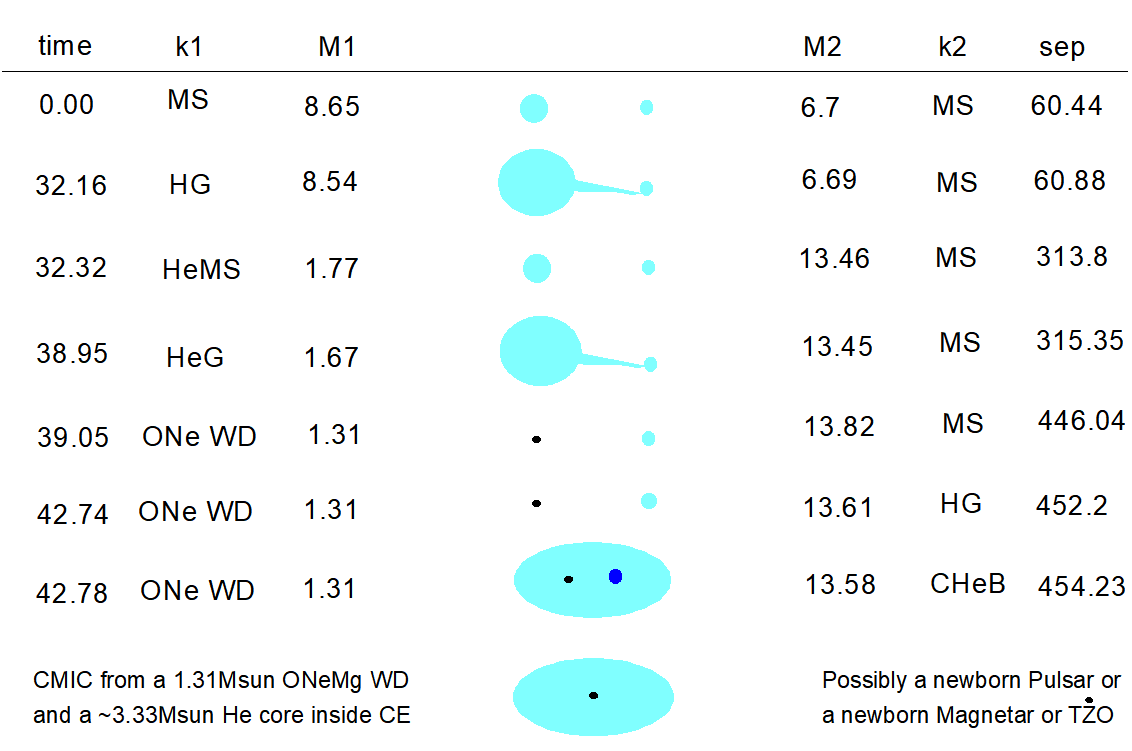}
\caption{Example of the evolution of a system to the channel for the merger between
a ONeMg WD and He core of a massive star based on model 1. The primary (mass is M1) experiences two phases of mass transfer before forming
the ONeMg WD. Afterwards, the secondary (mass is M2) fills its
Roche lobe at the core helium burning stage of the evolution (CHeB), and its envelope engulfs the ONeMg WD and its core due to the unstable mass transfer.
The abbreviations of the stellar types (k1 is the type of the primary star, and k2 is type of the secondary star in a binary) are defined in the text. The evolution time is in Myr, and separation (sep) between two stars is in $R_\odot$.}
\label{fig:1}
\end{figure*}

\clearpage

\begin{figure*}
\centering
\includegraphics[totalheight=2.4in,width=2.9in]{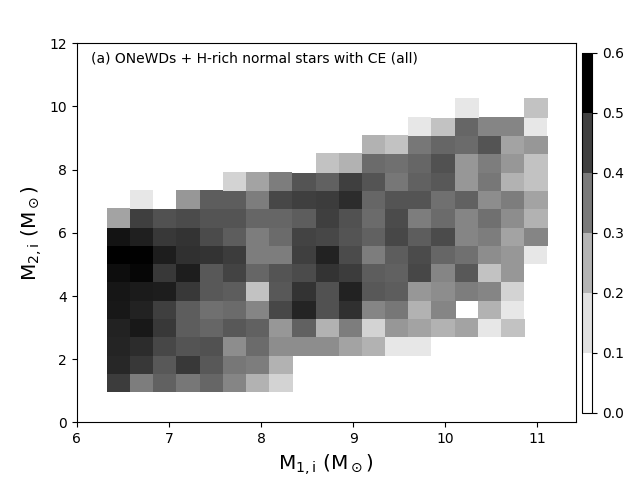}
\includegraphics[totalheight=2.4in,width=2.9in]{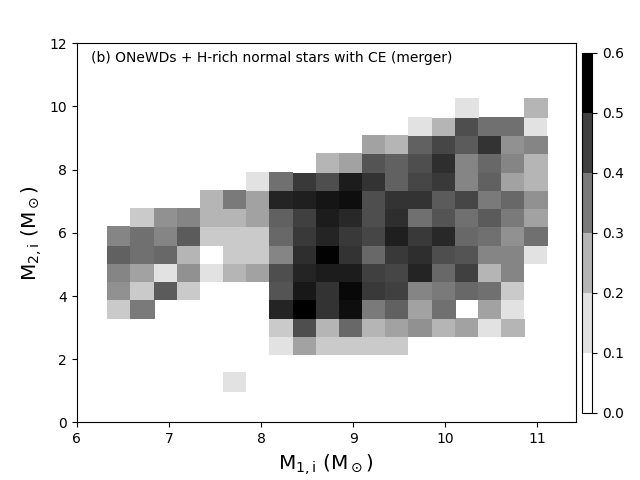}
\includegraphics[totalheight=2.4in,width=2.9in]{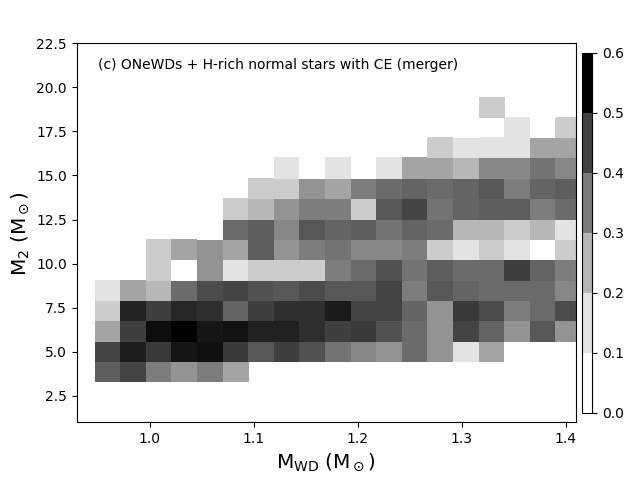}
\includegraphics[totalheight=2.4in,width=2.9in]{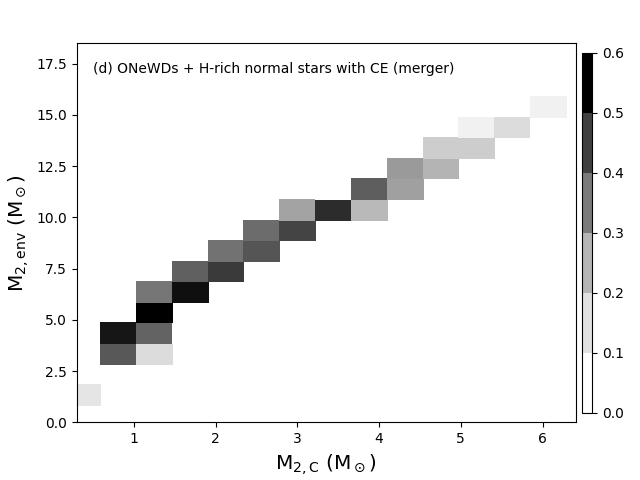}
\caption{The normalized/weighted number density distributions from simulated results of model 1 (standard model). (a) and (b) show the initial mass relation of primary MS ($M_{\rm 1,i}$) and secondary MS stars ($M_{\rm 2,i}$) in primordial binaries which form all ONeMg WD + normal star binaries with CE (upper left), and ONeMg WD + normal star systems which experience the CE and leading to core mergers (upper right), respectively. Mass relations at the beginning of the CE between the ONeMg WD and its companion star for ONeMg WD + normal star systems, and secondaries' core and envelope mass relations for the merger case under the condition of $M_{\rm WD}+M_{\rm 2,C}\geq1.44$ are given in (c) and (d), respectively. }
\label{fig:1}
\end{figure*}

\clearpage

\begin{figure*}
\centering
\includegraphics[totalheight=3.7in,width=4.8in]{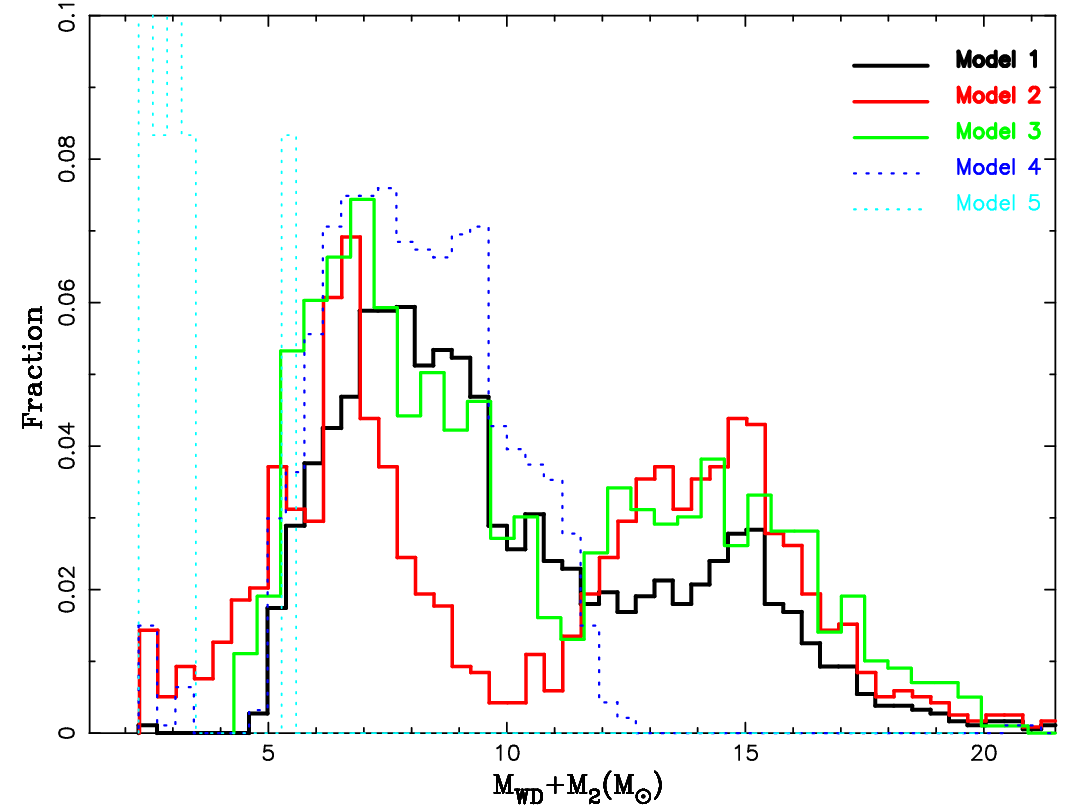}
\caption{The total mass distributions of merger systems between the ONeMg WDs and
cores of the normal stars under the condition of $M_{\rm WD}+M_{\rm 2,C}\geq1.44$ for five different models.}
\label{fig:1}
\end{figure*}

\clearpage

\begin{figure*}
\centering
\includegraphics[totalheight=2.6in,width=3.6in]{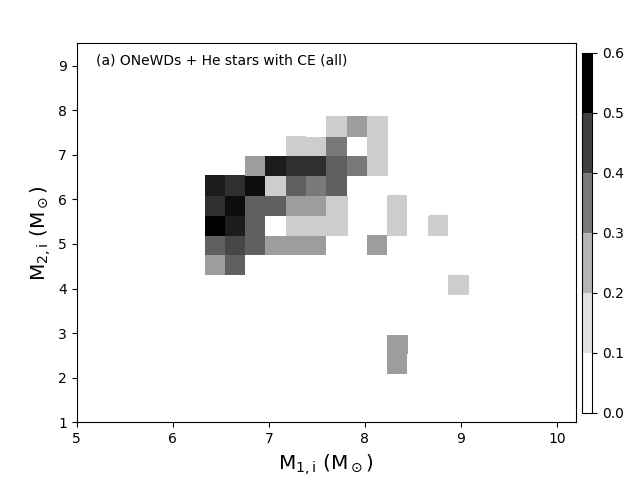}
\includegraphics[totalheight=2.6in,width=3.6in]{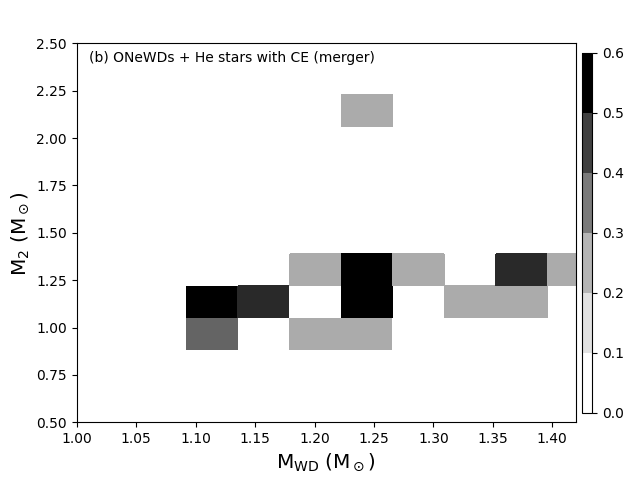}
\includegraphics[totalheight=2.6in,width=3.6in]{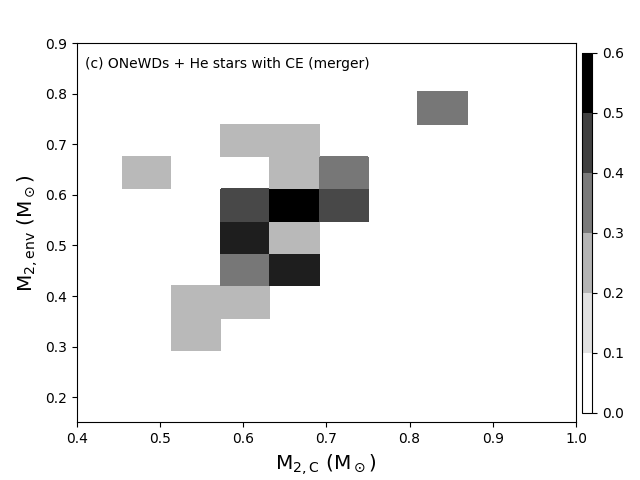}
\caption{The normalized/weighted number density distributions from simulated results of model 1 (standard model) for the merger case of the ONeMg WD and core of the He stars under the condition of $M_{\rm WD}+M_{\rm 2,C}\geq1.44$ at the onset of CE. (a) demonstrates the initial mass relation of primary MS ($M_{\rm 1,i}$) and secondary MS stars ($M_{\rm 2,i}$) in primordial binaries which form all ONeMg WD + He star binaries with CE. (b) shows the mass relation of the ONeMg WD ($M_{\rm WD}$) and secondary He stars ($M_{\rm He}$). Mass relations between the ONeMg WD and core of the He star for the merger case are given in (c).}
%\label{fig:1}
\end{figure*}

\clearpage

\begin{figure*}
\centering
\includegraphics[totalheight=3.7in,width=4.8in]{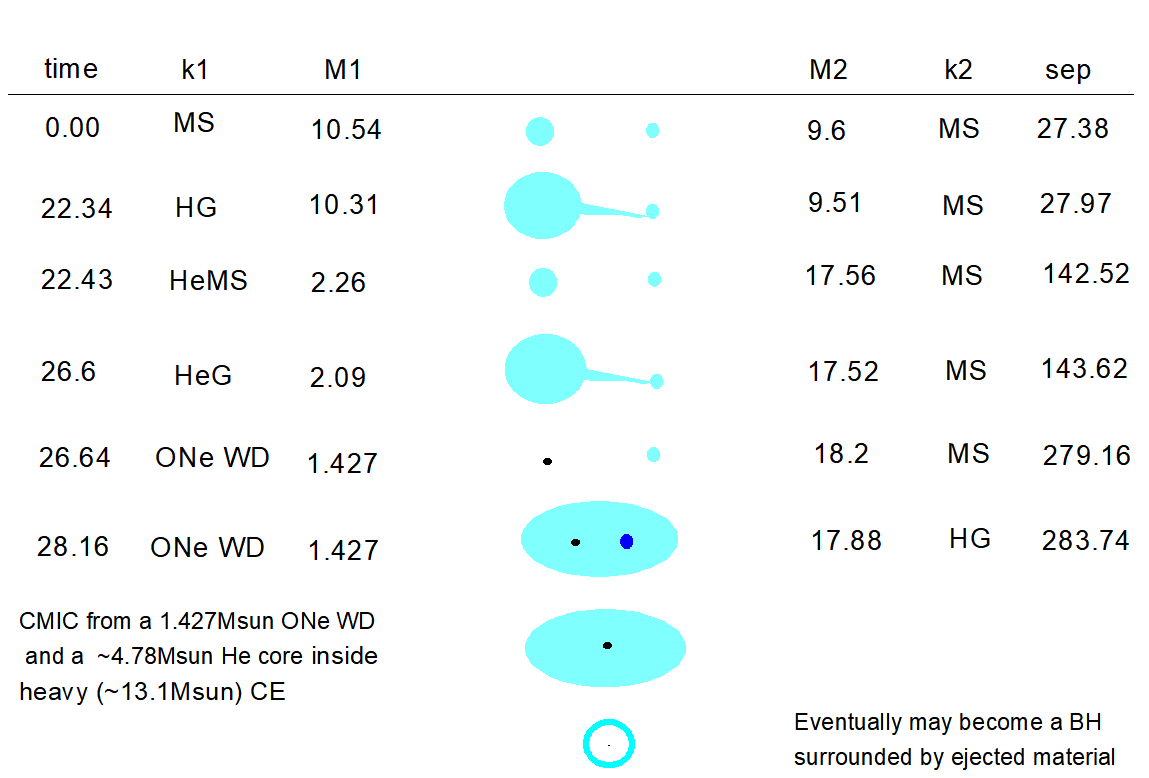}
\caption{Example of the merger between a ONeMg WD and He core of a massive star inside the CE based on model 1. The evolution is very similar to that of Fig. 1, and the main difference compared to the first example is that the combined core and total mass are higher, the CE event happens when the secondary becomes the HG star, and the evolution time is longer due to the different initial conditions. The column headings and units are same as that of Fig. 1.}
\label{fig:1}
\end{figure*}

\clearpage

%%%%%%%%%%%%%%%%%%%%%%%%%%%%%%%%%%%%%%%%%%%%%%%%%%

%%%%%%%%%%%%%%%%% APPENDICES %%%%%%%%%%%%%%%%%%%%%

%\appendix



%%%%%%%%%%%%%%%%%%%%%%%%%%%%%%%%%%%%%%%%%%%%%%%%%%

% Don't change these lines
%\bsp	% typesetting comment
\label{lastpage}
\end{document}